# Machine Learning Based Approach to Recommend MITRE ATT&CK Framework for Software Requirements and Design Specifications

Nicholas Lasky
Department of Computer Science
University of Wisconsin-Eau Claire
laskyn0042@uwec.edu

Benjamin Hallis
Department of Computer Science
University of Wisconsin-Eau Claire
hallisbj4335@uwec.edu

Dr. Mounika Vanamala
Department of Computer Science
University of Wisconsin-Eau Claire
vanamalm@uwec.edu

Dr. Rushit Dave
Department of Computer Science
Minnesota State University
rushit.dave@mnsu.edu

Dr. Jim Seliya
Department of Computer Science
University of Wisconsin-Eau Claire
seliyana@uwec.edu

## Abstract

*Engineering more secure software has become a critical challenge in the cyber world. It is very important to develop methodologies, techniques, and tools for developing secure software. To develop secure software, software developers need to think like an attacker through mining software repositories. These aim to analyze and understand the data repositories related to software development. The main goal is to use these software repositories to support the decision-making process of software development. There are different vulnerability databases like Common Weakness Enumeration (CWE), Common Vulnerabilities and Exposures database (CVE), and CAPEC. We utilized a database called MITRE. MITRE ATT&CK tactics and techniques have been used in various ways and methods, but tools for utilizing these tactics and techniques in the early stages of the software development life cycle (SDLC) are lacking. In this paper, we use machine learning algorithms to map requirements to the MITRE ATT&CK database and determine the accuracy of each mapping depending on the data split.*

**Keywords**: machine learning, MITRE, functional requirement, security, algorithm, training set, testing set, data split

## Introduction

There is a wide range of security attacks from DDoS to spear phishing (Melnick, 2018). It is the developers' job to minimize the risk of security breaches, but it is very hard to protect against every possibility. There are several ways to bypass security, and it is incredibly difficult to create a system that protects against every single one. For example, the financial sector contains lots of sensitive, personal data regarding names, social security numbers, money, and more. All this important data increases the risk of the data being breached. It is a necessity to cover all bases when designing systems for banks, and to do that, security must be a major focus at every stage of the project. Using machine learning to find and classify vulnerabilities is a promising method. Developers that are working on projects are required to spend a lot of time looking through the project and security attack databases to make sure there are no vulnerabilities. The problem with this is that there are tons of possible methods of attacks, and it is difficult for developers to find them all (Vanamala et al., 2020).

MITRE has developed MITRE ATT&CK, a public knowledge-based library of adversary tactics and techniques. MITRE has been used by several organizations in their products and processes ("MITRE

ATT&CK", 2022). MITRE has 14 branches which describe the type of attack potentially being carried out. Each branch has multiple subcategories where it goes into different techniques used by an attacker. The 14 branches are as follows: reconnaissance (10 techniques), resource development (7 techniques), initial access (9 techniques), execution (12 techniques), persistence (19 techniques), privilege escalation (13 techniques), defense evasion (40 techniques), credential access (15 techniques), discovery (29 techniques), lateral movement (9 techniques), collection (17 techniques), command and control (16 techniques), exfiltration (9 techniques), and impact (13 techniques). Reconnaissance is when an attacker gathers information for future operations. Resource development establishes resources to support operations. Initial access describes when someone is trying to break into your network. Execution involves trying to run malicious code. Persistence is when a user tries to maintain their foothold within a system. Privilege escalation seeks to try and gain higher permissions. Defense evasion aims to try and hide an attacker without detection. Credential access involves stealing names and passwords to get into a system. Discovery tries to figure out the environment and see what an attacker can control. Lateral movement is when an attacker moves throughout your environment to get through multiple systems. Collection is when an attacker gathers data to help them in their goals. Command and control involve an attacker communicating with compromised systems to take control of them. Exfiltration is when an attacker steals data. Finally, impact aims to manipulate and/or destroy systems and data. If you go into each technique in a branch, it provides different procedures/attacks carried out to find vulnerabilities in SRS, but it also provides mitigations and detections.

As cyber-attacks increase in volume and sophistication, the current state of cybersecurity solutions is inadequate. According to Red Canary, Advanced Persistent Threat (APT) attacks have increased from approximately 500 attacks per year in 2009 to almost 2,500 APT attacks per year in 2019 (Vanamala et al., 2020). The lack of timely detection and response is mainly caused by the insufficient support of attack action correlations and prediction to allow for proactive intrusion, investigation, and mitigation. There are lots of different ways to try and get past security, and it is difficult to create a system that protects against every single one (Vanamala et al., 2020). With the vast amount of data available online, it is necessary to integrate security-related activities and deliverables into each phase of the SDLC.

Although MITRE ATT&CK TTP has been used in various ways, none of the above uses help software developers utilize these techniques and tactics in developing secure software in the early stages of the SDLC, such as requirements and design. With the large amount of information provided in MITRE, it is difficult for software developers to go through ATT&CK TTP's manually and identify those relevant to the software they are developing. In this study, we use machine learning to map requirements to MITRE and determine the accuracy of each mapping.

## Related Work

In the Atlantis Press, an article was posted assessing the risk of security non-compliance of banking security requirements based on attack patterns. They analyzed the applications of information systems (IS) in banking to ensure that security regulations are in place (Rongrat and Senivongse, 2018). A risk index model was used, which determined the risk level based on how severe and how likely an attack pattern could affect the system if regulations aren't in place. For evaluation, they utilized F-measure and accuracy, which is a commonality that our research shares with theirs. For taxonomy, they utilized CAPEC to map their attack patterns while we used MITRE (Rongrat and Senivongse, 2018). In industry, it's difficult to reuse previously written software modules for banks, so instead, they utilized software requirements. They would match written requirements of the new software against requirements used to define the old software. When comparing, if there were requirement pairs between the new and old software, they would reuse these pairs. To measure degree of similarity, they used cosine, Dice, and Jaccard coefficients. To increase efficiency, they automated the integration of software requirements by using text similarity, which would find duplicate requirements to avoid doing the same job twice. To determine the security satisfaction of requirements, they used a Lucene-based search engine for attacks and weaknesses via CAPEC (Rongrat and Senivongse, 2018). The keywords from each requirement are gathered and searched for in CAPEC to find relevant attacks and weaknesses, and this is how the risk level is assessed. For example, one standard security requirement of banking (SSRB) could be "The system shall enforce strong password (contain a mix of alphabetic and non-alphabetic characters)." Then, in CAPEC, their tool gathered keywords from the given SSRB and mapped it to two possible attack patterns: CAPEC_ID 16 and CAPEC_ID 49. CAPEC_ID 16 is made for creating a strong password policy and ensure that your system enforces this policy, and CAPEC_ID 49 is made to put together a strong password policy and make sure that all user created passwords comply with it. Both CAPEC IDs are very similar but also unique, and their tool successfully mapped the SSRB to several possible attack patterns, along with their mitigations. Then, they used SSRBs and banking security requirements (SRBs), and they wanted to see how compliant the SRBs are to the SSRBs. Text similarity was used here once again. For their risk index, they calculated it by taking the product of the probability of a risk and the severity of impact caused by the risk. The probability of risk of attacks is associated with an SSRB when it is not met by any SRBs of a bank. They did this to see which SSRBs were missing from the SRBs of the banks. In terms of our research, we are looking at broad spectrum functional requirements, and we are utilizing a different taxonomy as mentioned above (MITRE instead of CAPEC). In terms of methodology, they used text similarity to see how compliant requirements were, but we have adopted a machine learning approach because of potential inaccuracies with text similarity in the mapping to a MITRE attack pattern. Overall, we want to evaluate human measures compared to our machine learning approach, and if both matches, our approach can be used to develop a tool for businesses to use in the real world to implement security requirements instead of having to spend valuable time and resources on this process.

In our research, we focused on gathering functional requirements and seeing how they relate to the MITRE ATT&CK taxonomy ("Functional vs Non Functional Requirements", 2020). We accomplished this through machine learning algorithms, such as random forest, neural networks, SVM, and Naive Bayes, which were then evaluated for accuracy. To train our system, we individually mapped each functional requirement to a potential MITRE branch. Once the mapping was done, we used this to see how accurate our training trials were. Possible error could include human error in the mapping to MITRE branches. In terms of future research, we want to apply a similar approach to develop a tool for industry to use, most likely in the financial sector, to successfully implement security requirements for companies automatically to reduce the overhead of a person manually mapping requirements to attack patterns.

## Flow Diagram

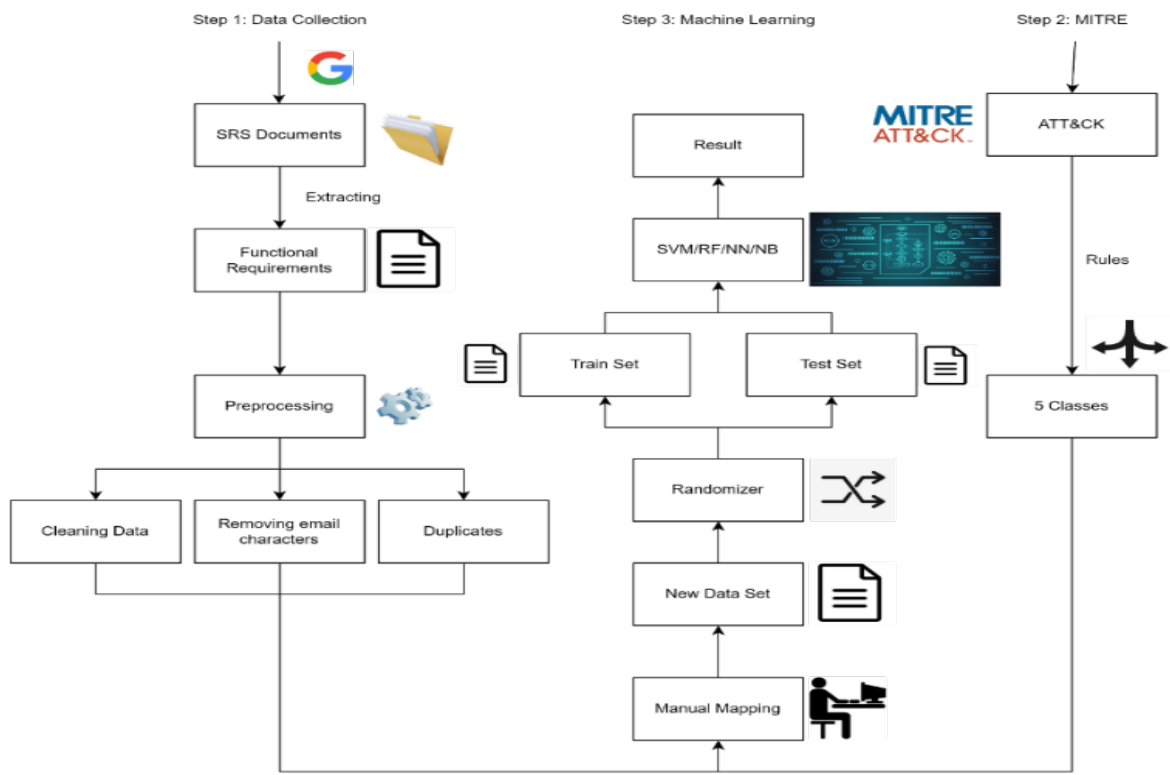

**Figure 1.  Research Process Model**

## Methodology

### *Phase 1 – Data Collection*

We collected functional requirements from multiple SRS documents. We accomplished this through Google searches and GitHub to find databases and documents with requirements. We obtained 614 unique functional requirements, which we then manually mapped to the MITRE taxonomy database, which determines potential attacks related to each requirement.

### *Phase 2 – Manual Mapping of Functional Requirements to MITRE*

We related each functional requirement to a technique within a branch, and once we determined a connection, we wrote down each branch as a class to use for our machine learning algorithms. However, due to too many classes influencing accuracy, we modified our class system and grouped certain branches together. We utilized 12 of the 14 MITRE branches for our classes. We came up with five total classes: C1: initial access, execution, and impact; C2: resource development and command and control; C3: persistence and defense evasion; C4: privilege escalation, credential access, and lateral movement; C5: collection and exfiltration. This system allowed the machine learning algorithms to more accurately predict where each individual requirement would map to one of the five classes.

The manual mapping was done in Excel by relating the functional requirements, written in plain English sentences, to the MITRE taxonomy, and once formatted correctly, we converted the Excel sheet to a CSV file, which then was applied in our machine learning algorithms to train itself on how to map each requirement to a potential class. The mapping relied on the class structure that we developed above. Two people manually mapped the requirements, so depending on how each person interpreted a requirement, it may change the way requirements were mapped from person to person. Each person was responsible for mapping half of the functional requirements dataset.

### *Phase 3 – Machine Learning Algorithms*

We applied four machine learning algorithms: neural networks, SVM, random forest, and Naive Bayes. A neural network uses a series of algorithms that tries to find relationships in a set of data through processes like the way the human brain works (Ray, 2017). SVM (Support Vector Machine) is a classification algorithm that plots each data item as a point in n-dimensional space with the value of each feature being the value of a coordinate, where n represents the number of features (Ray, 2017). Random forest is an ensemble of decision trees, and each tree gives a classification and votes for a class. Random forest will then choose the classification having the most votes among all the trees in the forest. Naive Bayes is a classification algorithm based on Bayes' Theorem where each predictor is independent of one another. The algorithm calculates the probability of an event given another event, typically represented as $P(A|B) = (P(B|A) * P(A)) / P(B)$, where A and B are independent events (Ray, 2017).

We developed our own randomizer program, which would randomly pick which requirements would be trained or tested. Our randomizer used the package pandas for reading in the original data file, and the sklearn package for splitting the data into two parts ("Scikit-Learn In Python", 2022). The randomizer uses pandas to read in a CSV file that holds all the sorted data and puts it into a data frame ("Package Overview", 2022). The user enters what percentage of the data they want to go into the training set. The sk.model_selection.train_test_split function uses that percentage to put data into the training set, and the remaining data after that into the testing set ("Scikit-Learn In Python", 2022). There are a couple lines that reset the index numbers for the data to make sure they aren't still being read in the same order as before. The training and testing sets are sent to their respective CSV files correctly sorted, so the other programs can easily pull from the updated CSV files. To run our data, we randomized the data sets and ran them through an algorithm. We repeated this process for the other three algorithms as well. We focused on which algorithm provided the best performance by looking at accuracy, recall, F-1 score, and precision (Harikrishnan, 2019).

# Results and Discussions

After training and testing our data, we concluded that all four algorithms performed the best with a 70% training and 30% testing data split, with SVM performing the best on average between all splits, followed closely by Naive Bayes, neural network, and random forest. The gaps between the other splits weren't noticeably different, which is primarily due to the rather small data set. Another factor that influenced the accuracy of each algorithm was the manual mapping, which was done by multiple people. Each person may have different thought processes on how to categorize functional requirements, and a few of them were difficult to put into a class because they didn't fit a certain "mold" or class. To up these accuracy measures in the future, we would try to obtain more functional requirements so that the machine would get better and better at training itself by providing it with a vast amount of data. In addition, we would also try other algorithms to see if they could outperform the three algorithms used in this study. Eventually, with high enough accuracy measures, we could craft a tool for companies to put their functional requirements into automatically to show the mappings, potential threats, and mitigations to make their functional requirements even more secure.

| 60% Training | Accuracy | Recall | F-1 Score | Precision |
|---|---|---|---|---|
| Neural Network | .47 | .47 | .47 | .47 |
| Random Forest | .47 | .47 | .47 | .47 |
| Support Vector Machine | .47 | .47 | .47 | .47 |
| Naive Bayes | .49 | .49 | .49 | .49 |

**Table 1. 60% Training, 40% Testing Set Results**

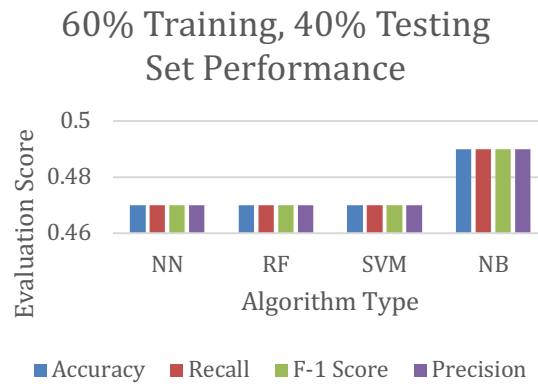

**Figure 2. Visual Results For 60% Training, 40% Testing Set**

| 70% Training | Accuracy | Recall | F-1 Score | Precision |
|---|---|---|---|---|
| Neural Network | .53 | .53 | .53 | .53 |
| Random Forest | .5 | .5 | .5 | .5 |
| Support Vector Machine | .57 | .57 | .57 | .57 |
| Naive Bayes | .51 | .51 | .51 | .51 |

**Table 2. 70% Training, 30% Testing Set Results**

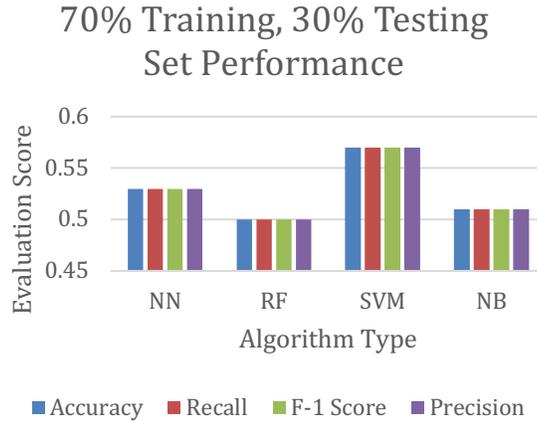

**Figure 3. Visual Results For 70% Training, 30% Testing Set**

| 80% Training | Accuracy | Recall | F-1 Score | Precision |
|---|---|---|---|---|
| Neural Network | .48 | .48 | .48 | .48 |
| Random Forest | .47 | .47 | .47 | .47 |
| Support Vector Machine | .48 | .48 | .48 | .48 |
| Naive Bayes | .5 | .5 | .5 | .5 |

**Table 3. 80% Training, 20% Testing Set Results**

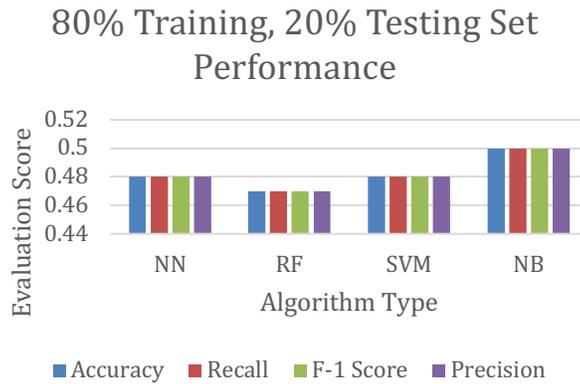

**Figure 4. Visual Results For 80% Training, 20% Testing Set**

After training and testing our data, we concluded that all four algorithms performed the best with a 70% training and 30% testing data split, with SVM performing the best on average between all splits, followed closely by Naive Bayes, neural network, and random forest. The gaps between the other splits weren't noticeably different, which is primarily due to the rather small data set. Another factor that influenced the accuracy of each algorithm was the manual mapping, which was done by multiple people. Each person may have different thought processes on how to categorize functional requirements, and a few of them were difficult to put into a class because they didn't fit a certain "mold" or class. To up these accuracy measures in the future, we would try to obtain more functional requirements so that the machine would get better and better at training itself by providing it with a vast amount of data. In addition, we would also try other algorithms to see if they could outperform the three algorithms used in this study. Eventually, with high enough accuracy measures, we could craft a tool for companies to put their functional requirements into automatically to show the mappings, potential threats, and mitigations to make their functional requirements even more secure.

## Conclusion and Future Work

We collected 614 functional requirements and manually mapped them to MITRE's database and applied four machine learning algorithms to train our system. We determined the accuracy of each algorithm's ability to recommend the correct attack from MITRE's database. We tested our data with four algorithms: neural networks, SVM, random forest, and Naive Bayes. Of the four algorithms, SVM performed the best on average with a 70% training set and 30% testing set data split. Overall, the accuracy gaps between data splits were not noticeably different due to the relatively small data set of 614 requirements. The accuracy of each algorithm was influenced by the manual mapping, which was done by multiple people, so each person will have a different thought process on how to map requirements. For the future, we need to obtain more functional requirements so that the machine can get better at training itself. In addition, we plan to incorporate more algorithms to train and evaluate sets to see which one offers the best performance. After consistent, high accuracies, we will craft a tool for companies to put their functional requirements in to automatically show the mappings, potential threats, and mitigations to make their software even more secure.